\documentclass[10pt,aps,twocolumn,nobibnotes,prl,showpacs]{revtex4-1}
  \usepackage{color}
  \usepackage{newlfont}
  \usepackage{graphicx}
  \usepackage{amssymb}
  \usepackage{amsmath}
  \usepackage{verbatim}
  \usepackage[latin1]{inputenc}
  \usepackage{bbm}
\newcommand{\ket}[1]{\mbox{$ | #1 \rangle $}}
\newcommand{\bra}[1]{\mbox{$ \langle #1 | $}}
\newcommand{\be}{\begin{eqnarray}}
\newcommand{\ee}{\end{eqnarray}}
\begin{document}
\title{A measure of physical reality}
\author{A. L. O. Bilobran}
\author{R. M. Angelo}
\affiliation{Department of Physics, Federal University of Paran\'a, P.O.Box 19044, 81531-980, Curitiba, PR, Brazil}
\begin{abstract}
From the premise that an observable is real after it is measured, we envisage a tomography-based protocol that allows us to propose a quantifier for the degree of indefiniteness of an observable given a quantum state. Then, we find that the reality of an observable can be inferred locally only if this observable is not quantumly correlated with an informer. In other words, quantum discord precludes Einstein's notion of separable realities. Also, by monitoring changes in the local reality upon measurements on a remote system, we are led to define a measure of nonlocality. Proved upper bounded by discordlike correlations and requiring indefiniteness as a basic ingredient, our measure signals nonlocality even for separable states, thus revealing nonlocal aspects that are not captured by Bell inequality violations.
\end{abstract}

\maketitle

%==========================================
{\em Introduction}. From Descartes' dictum {\em cogito ergo sum} one might be tempted to conclude that a brain-endowed system living in an empty universe could ensure its own reality. This position, however, cannot be maintained within a scientific framework. The reason is that physics is relative by essence, so it is not possible for an object to ascribe any physical state to itself; a reference frame is needed. It follows, therefore, that any attempt to build an empirically accessible notion of physical reality demands in the first place the definition of two entities, namely, an ``observer'' and an ``observed'', whose roles are interchangeable. These objects can interact and get to know about each other, in which case it then makes sense for one to speak of the physical reality of the other. 

This is precisely what we have in ordinary situations. Our notion of reality is nurtured by the process of repeated measurements that takes place, e.g., every time we watch an object at rest. The huge amount of photons that reach our retina after being scattered by the object---without appreciably disturbing it---informs us that the object ``exists'' in a ``definite'' position, so it is {\em real}. Although the sensation of reality is granted to the person who receives the scattered photons, the information about the object was already encoded in the photons. Such information, which manifests as correlations in the system ``object + photons'', was generated via physical interactions which by no means depend on the very existence of a retina or a brain. Hence, the capability of a system of informing the presence of another is a primordial condition for the existence of some element of physical reality. Moreover, without such a mechanism, reality cannot be empirically probed.

In 1935, an attempt was put forward by Einstein, Podolsky, and Rosen (EPR)~\cite{EPR} aiming at defining the notion of {\em element of reality}: ``If, without in any way disturbing a system, we can predict with certainty the value of a physical quantity, then there exists an element of physical reality corresponding to this physical quantity.'' Along with the conception that ``every element of the physical reality must have a counterpart'' in a {\em complete theory}, this definition immediately implies that either ``(A) the quantum-mechanical description of reality given by the wavefunction is not complete or (B) when the operators corresponding to two physical quantities do not commute the two quantities cannot have simultaneous reality.'' By tacitly assuming locality and arguing that there are quantum states for which noncommuting observables can be simultaneously real, EPR proved (B) wrong and claimed the incompleteness of quantum theory. Later on, Bell showed that any theory aiming at completing quantum mechanics would be unavoidably nonlocal~\cite{bell}, Bohmian mechanics~\cite{bohm} being a prominent illustration of that. Other approaches appeared defending purely statistical interpretations for the quantum formalism~\cite{ballentine}, in consonance with Einstein's view. 

Discussions about foundational aspects of quantum theory, particularly on the wave function interpretation, entered the 21st century with physicists polarized in two main lines of thought, both supported by substantial amount of theoretical work. While on one hand $\psi$-ontic models ascribe to the wave function a realistic nature, on the other $\psi$-epistemic models suggest that it actually is knowledge about an underlying reality. Recent years have witnessed significant contributions in favor of $\psi$-ontic models~\cite{PBR,rudolph12,renner12,hardy13,massar13,lowther13,leifer14,maroney14,branciard14} against a more modest number of works exploring $\psi$-epistemic ones~\cite{spekkens07,spekkens10,veitch13,farr14}. Although general and powerful for their purposes, these works offer no clear interpretation for mixed states. In a different vein, considerable progress has recently been made towards a deep understanding of the quantum measurement problem and the emergence of objective classicality in the framework of the {\em quantum Darwinism}~\cite{zurek09,horodecki15}.

Here, we aim at discussing the notion of physical reality by focusing not only on the quantum state but also on observables and their measurements. In particular, in what follows we will formulate a quantifier of reality that is grounded on an empirical protocol. Then, some relevant implications are presented, in particular a notion of nonlocality that is not captured by Bell inequality violations. We close the paper with a summary of our results and a brief remark positioning our proposal in the context delineated by EPR's and Bell's works.

\begin{figure}[t]
\centerline{\includegraphics[scale=0.1]{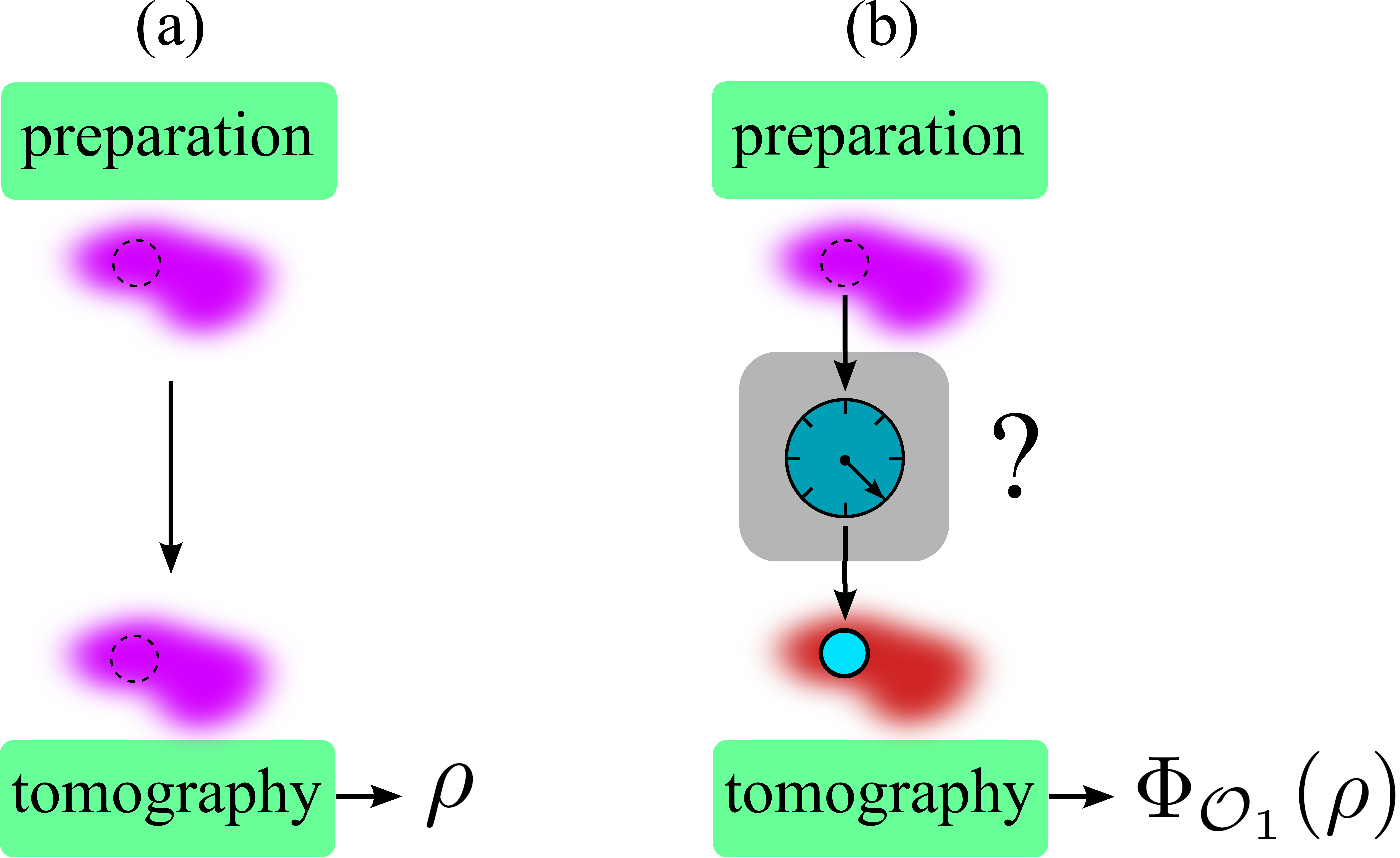}}
\caption{(Color online). (a) A preparation $\rho$ is determined by state tomography. (b) An observable $\mathcal{O}_1$ is secretly measured after the preparation, so that it is surely real before the tomography, which then predicts a state $\Phi_{\mathcal{O}_1}(\rho)$. If $\Phi_{\mathcal{O}_1}(\rho)=\rho$, then the secret measurement has just revealed a pre-existing element of reality.}
\label{fig1}
\end{figure}

%============================
{\em Elements of reality}. Consider an experimental procedure that prepares a physical state for a multipartite system. A task is defined which consists of determining, via state tomography, the most complete description for this preparation. We are allowed to repeat the procedure as many times as necessary to get an ideal tomography. Thus, at the end of the day, we get to know that, every time the procedure runs, the quantum mechanical description for the system will be, say, $\rho$ (Fig.~\ref{fig1}a). Then, we are exposed to a different scheme (Fig.~\ref{fig1}b). Again, we are asked to propose a complete description for the system state, given the same preparation and tomography process, but now a measurement of an observable $\mathcal{O}_1=\sum_ko_{1k}\mathrm{O}_{1k}$, with projectors $\mathrm{O}_{1k}=\ket{o_{1k}}\bra{o_{1k}}$ acting on $\mathcal{H}_1$, is secretly performed by an agent in between the preparation and the tomography, in every run of the procedure. Quantum theory predicts that the system will be in the state $\mathrm{O}_{1k}\otimes\rho_{2|o_{1k}}$ with probability $p_{o_{1k}}$ after the measurement is performed, where $\rho_{2|o_{1k}}=\text{Tr}_1(\mathrm{O}_{1k}\,\rho\,\mathrm{O}_{1k})/p_{o_{1k}}$ is the state of the rest of the system given the outcome $o_{1k}$ and $p_{o_{1k}}=\text{Tr}(\mathrm{O}_{1k}\rho\,\mathrm{O}_{1k})$ is the probability associated with this particular outcome. Without any information about agent's measurements, after the state tomography our best description will be
\be\label{Phi(rho)}
\Phi_{\mathcal{O}_1}(\rho)\equiv\sum_k\mathrm{O}_{1k}\,\rho\,\mathrm{O}_{1k}=\sum_kp_{o_{1k}}\mathrm{O}_{1k}\otimes\rho_{2|o_{1k}}.
\ee
Now, the agent is certain, by EPR's criterion, that the observable is real after each measurement is made. It follows, therefore, that our description \eqref{Phi(rho)} is {\em epistemic} with respect to $\mathcal{O}_1$, i.e., $p_{o_{1k}}$ reflects only our subjective ignorance about the actual value of $\mathcal{O}_1$.
The protocol proceeds with the comparison of the descriptions obtained in (a) and (b). When $\Phi_{\mathcal{O}_1}(\rho)=\rho$, the situation is such that the agent can conclude that an element of reality for $\mathcal{O}_1$ was implied by the very preparation. In this case, the agent measurements did not create reality, but revealed a pre-existing one. This suggests the following criteria of reality.
\vskip2mm
\noindent{\bf Definition} (Element of reality). {\em An observable $\mathcal{O}_1=\sum_k o_{1k}\mathrm{O}_{1k}$, with projectors $\mathrm{O}_{1k}=\ket{o_{1k}}\bra{o_{1k}}$ acting on $\mathcal{H}_1$, is real for a preparation $\rho\in\bigotimes_{i=1}^N\mathcal{H}_i$ if and only if}
\be\label{def1}
\Phi_{\mathcal{O}_1}(\rho)=\rho.
\ee
\vskip2mm

The map $\Phi_{\mathcal{O}_1}$ defined by Eq.~\eqref{Phi(rho)} denotes a procedure of unread measurements, as delineated by our protocol. Clearly, the criterion \eqref{def1} agrees with EPR's on the reality of $\mathcal{O}_1$ when the preparation is an eigenstate of this observable, i.e., $\rho=\mathrm{O}_{1k}$. But it also predicts an element of reality for a mixture of eigenstates, $\rho=\sum_kp_{o_{1k}}\mathrm{O}_{1k}$, as in this case $\Phi_{\mathcal{O}_1}(\rho)=\rho$.  Another interesting point is that the criterion \eqref{def1} automatically incorporates the fact that a measurement preserves a pre-existing reality, i.e., $\Phi_{\mathcal{O}_1\mathcal{O}_1}(\rho)\equiv\Phi_{\mathcal{O}_1}(\Phi_{\mathcal{O}_1}(\rho))=\Phi_{\mathcal{O}_1}(\rho)$. Finally, since a state in the form $\Phi_{\mathcal{O}_1}(\rho)$ can be viewed as an epistemic state with respect to $\mathcal{O}_1$, its collapse upon an eventual measurement of $\mathcal{O}_1$ can be interpreted as mere information updating rather than a physical process.

The above criteria naturally induces a measure of by how much a given state $\rho$ is far from a state with $\mathcal{O}_1$ real. We define the {\em indefiniteness} (or {\em irreality}) of the observable $\mathcal{O}_1$ given the preparation $\rho\in\mathcal{H}$ as the entropic distance
\be\label{frakI}
\mathfrak{I}(\mathcal{O}_1|\rho)\equiv S(\Phi_{\mathcal{O}_1}(\rho))-S(\rho),
\ee
where $S(\rho)$ is the von Neumann entropy. Because projective measurements can never reduce the entropy~\cite{chuang}, one has that $\mathfrak{I}(\mathcal{O}_1|\rho)\geq 0$. Since the von Neumann entropy is strictly concave, it follows that $\mathfrak{I}$ will be zero (i.e., $\mathcal{O}_1$ will be real) if and only if the condition \eqref{def1} holds~(see Ref.~\cite{ana13} for a related demonstration).

%=====================
{\em Implications}. The question then arises as to whether the measure \eqref{frakI} can furnish insights to further aspects of quantum theory. To start with, we invoke the Stinespring theorem~\cite{spehner14}, $\Phi(\rho)=\text{Tr}_A\left(U\rho\otimes\ket{a_0}\bra{a_0}U^{\dag}\right)$, according to which any quantum operation $\Phi$ can be viewed as a reduced evolution of the system coupled to an ancillary system $\mathcal{A}$, where $U$ is a unitary operator acting on $\mathcal{H}\otimes\mathcal{H}_{\mathcal{A}}$ and $\ket{a_0}\in\mathcal{H}_{\mathcal{A}}$. This observation suggests that reality emerges upon the dynamical generation of correlations between the system and some {\em informer}, i.e., a degree of freedom that records the information about the physical state of the system and is discarded (or secretly read, as in the protocol of Fig.~\ref{fig1}). This point is illustrated by Bohr's floating-slit experiment~\cite{bohr} (see Ref.~\cite{miron15} for a recent realization of Bohr's thought experiment). After interacting with a light floating slit S, a particle P moves towards a double-slit system, which is rigidly attached to the laboratory. Momentum conservation implies that in order for P to move towards the upper (lower) slit, S has to move downwards (upwards). If $m$ and $M$ denote the masses of P and S, respectively, then the correlation generated in this experiment can be described by the state $\ket{\Psi}=\tfrac{1}{\sqrt{2}}(\ket{v}_{\text{\tiny P}}\ket{-\tfrac{mv}{M}}+\ket{-v}_{\text{\tiny P}}\ket{\tfrac{mv}{M}})$, where $v$ and $mv/M$---the speeds of P and S, respectively---are treated as discrete variables, for simplicity. It is just an exercise to show that $\mathfrak{I}(\mathrm{v}|\rho_{\text{\tiny P}})$, with $\rho_{\text{\tiny P}}=\text{Tr}_{\text{\tiny S}}\ket{\Psi}\bra{\Psi}$, is a monotonically increasing function of $x=|\langle\tfrac{mv}{M}|-\tfrac{mv}{M}\rangle|$ and that $\mathfrak{I}(\mathrm{v}|\rho_{\text{\tiny P}})=0$ only if $x=0$. This shows that the velocity $\mathrm{v}$ of P given $\rho_{\text{\tiny P}}$ will be real only if the motion of S can be unambiguously identified, i.e., if the slit can properly play the role of an informer, in which case no interference pattern will be seen. Clearly, the reality of the velocity can be adjusted by the ratio $\tfrac{m}{M}$, whose value is previously chosen by the observer (in consonance with Bohr's view~\cite{bohr}). When $m\ll M$ (a nearly fixed slit), the momentum conservation will not be able to reveal the path of the particle, so the velocity will be maximally indefinite and interference fringes will appear.

%==================================
{\em Reality inseparability}. Consider the indefiniteness $\mathfrak{I}(\mathcal{O}_1|\rho)$ of an observable $\mathcal{O}_1$ given a preparation $\rho$. It is straightforward to check that
\be\label{IO1}
\mathfrak{I}(\mathcal{O}_1|\rho)=\mathfrak{I}(\mathcal{O}_1|\rho_1)+D_{[\mathcal{O}_1]}(\rho),
\ee
where $\rho_1=\text{Tr}_2\rho$ and $D_{[\mathcal{O}_1]}(\rho)=I_{1:2}(\rho)-I_{1:2}(\Phi_{\mathcal{O}_1}(\rho))$. Up to a minimization, $D_{[\mathcal{O}_1]}$ is a discord-like measure~\cite{ollivier01,henderson01,rulli11} written in terms of the mutual information $I_{j:k}(\rho)=S(\rho||\rho_j\otimes\rho_k)$ of the parts $j$ and $k$, where $S(\rho||\sigma)=\text{Tr}(\rho\ln\rho-\rho\ln\sigma)$ is the relative entropy. $\mathfrak{I}(\mathcal{O}_1|\rho_1)$ can be viewed as a measure of {\em local indefiniteness}, as it quantifies the indefiniteness of $\mathcal{O}_1$ given the local state $\rho_1=\text{Tr}_2\rho$. This quantity has recently been used to quantify waviness and coherence~\cite{angelo15,plenio13}. The relation \eqref{IO1} can be rewritten as
\be \label{D1}
\mathfrak{I}(\mathcal{O}_1|\rho)-\mathfrak{I}(\mathcal{O}_1|\rho_1)\geq \mathcal{D}_1(\rho),
\ee
where $\mathcal{D}_1(\rho)=\min_{\mathcal{O}_1}D_{[\mathcal{O}_1]}(\rho)$ is the quantum discord. Interestingly, this shows that an amount $\mathcal{D}_1$ of quantum correlation prevents the indefiniteness of $\mathcal{O}_1$ given the tomography $\rho$ to be equal to its indefiniteness given the local tomography $\rho_1$. Meaning that the reality of $\mathcal{O}_1$ cannot be devised separately from the other subsystems, even when they are far apart, this constitutes a violation of Einstein's separability principle~\cite{howard85}. It is worth noticing that, in a related context, Wiseman has recently identified the fundamental role of quantum discord in defining Bohr's notion of disturbance in the Bohr-EPR debate~\cite{wiseman13}.

Turning to the floating-slit experiment, when the slit is light enough we have that $\mathfrak{I}(\mathrm{v}|\rho_{\text{\tiny P}})=0$, so $\mathrm{v}$ is {\em locally definite} and, accordingly, no interference fringe will be observed. On the other hand, $\mathfrak{I}(\mathrm{v}|\ket{\Psi})=\mathcal{D}_{\text{\tiny P}}(\ket{\Psi})=\ln 2$ (the amount of entanglement in $\ket{\Psi}$), so $\mathrm{v}$ is {\em globally indefinite}. After all, is there an element of reality associated with $\mathrm{v}$? The answer depends on how the reality is probed. In an interference experiment, only the particle is monitored, so that it is the local reality that is accessed. If we look at both the particle and the slit, then we will be able to identify correlations, which will blur our inference about the reality of the particle.

%=================================================
{\em Irreality of incompatible observables}. Consider a preparation $\rho$ for a multipartite system and two mutually unbiased bases (MUBs), $\{\ket{o_{1k}}\}$ and $\{\ket{o'_{1k}}\}$ in $\mathcal{H}_1$, associated with maximally incompatible observables (MIO) $\mathcal{O}_1$ and $\mathcal{O}'_1$, respectively. Let us compute $\mathfrak{I}(\mathcal{O}'_1|\Phi_{\mathcal{O}_1}(\rho))$, i.e., the irreality of $\mathcal{O}'_1$ given a state $\Phi_{\mathcal{O}_1}(\rho)$ with $\mathcal{O}_1$ real. Since $|\langle o_{1k}|o'_{1k'}\rangle|^2=\tfrac{1}{d_1}$ for MUBs, where $d_1=\dim\mathcal(\mathcal{H}_1)$, one shows that $\Phi_{\mathcal{O}'_1\mathcal{O}_1}(\rho)=\tfrac{\mathbbm{1}_1}{d_1}\otimes \rho_2$, where $\rho_2=\text{Tr}_1\rho$. It follows that $\mathfrak{I}(\mathcal{O}'_1|\Phi_{\mathcal{O}_1}(\rho))+\mathfrak{I}(\mathcal{O}_1|\rho)=I_{1:2}(\rho)+\mathcal{I}(\rho_1)$, where $\mathcal{I}(\rho_1)\equiv \ln d_1-S(\rho_1)\geq 0$ is the available information~\cite{costa14}. Now, consider a preparation for which $\mathcal{O}_1$ is real, i.e., $\rho=\Phi_{\mathcal{O}_1}(\rho)$. Then, $\mathfrak{I}(\mathcal{O}_1|\rho)=0$ and
\be \label{Ixz>I1:2}
\mathfrak{I}(\mathcal{O}'_1|\Phi_{\mathcal{O}_1}(\rho))=I_{1:2}(\Phi_{\mathcal{O}_1}(\rho))+\mathcal{I}(\Phi_{\mathcal{O}_1}(\rho_1)).
\ee 
Hence, two MIO will be simultaneously real only if both terms in the right-hand side vanish. This will be the case only if $\rho=\tfrac{\mathbbm{1}_1}{d_1}\otimes\rho_2$, which is a fully uncorrelated state with a maximally incoherent reduced state. In this circumstance, all observables acting on $\mathcal{H}_1$ are simultaneously real, which renders to $\rho$ a fully classical essence. Clearly, this is not so for either an entangled state (for which $I_{1:2}>0$) or a projectively measured state (for which $\mathcal{I}>0$). Thus, in contrast with EPR's view, our notion of reality validates alternative (B) (see the Introduction), namely, that incompatible observables cannot be simultaneously real in general. 

%=======================
{\em Nonlocality}. Given two spacelike separated systems, can a physical action on one influence the reality of the other? Let us consider a typical EPR scenario, in which two subsystems, 1 and 2, prepared in a state $\rho$, are sent to distinct laboratories separated by a distance $d$. An ancillary system $A$, prepared in the state $\rho_A=\ket{a_0}\bra{a_0}$, is allowed to locally interact with the subsystem 2 through a unitary transformation $U_{2A}(t)$. Let $\tau$ be the time interval necessary for the completion of the injective information storage $U_{2A}(\tau)\ket{o_{2k}}\ket{a_0}=\ket{o_{2k}}\ket{a_k}$, where $\{\ket{a_k}\}$ is an orthonormal basis in the space $\mathcal{H}_{\mathrm{A}}$ and $\{\ket{o_{2k}}\}$ an orthonormal basis in $\mathcal{H}_2$. After the interaction ceases, the global state reads $\varrho(\tau)=U_{2A}(\tau)\rho\otimes\rho_AU_{2A}^{\dag}(\tau)$. The assumption of spacelike separated systems demands that $d\gg c\tau$. By focusing on the time evolution of the system of interest, we consider the following measure of nonlocality:
\be \label{NO1UA}
\mathcal{N}(\mathcal{O}_1,U_{2A}|\rho)\equiv \mathfrak{I}(\mathcal{O}_1|\text{Tr}_A\varrho(0))-\mathfrak{I}(\mathcal{O}_1|\text{Tr}_A\varrho(\tau)).
\ee
Clearly, this measure can signal alterations in the reality of $\mathcal{O}_1$ after the remote subsystem 2 has suffered a local perturbation. It is not difficult to show, from the above assumptions, that $\text{Tr}_A\varrho(\tau)=\Phi_{\mathcal{O}_2}(\rho)$, where $\mathcal{O}_2=\sum_ko_{2k}\ket{o_{2k}}\bra{o_{2k}}$. Thus, we arrive at
\be \label{NO1O2}
\mathcal{N}(\mathcal{O}_1,\mathcal{O}_2|\rho)=\mathfrak{I}(\mathcal{O}_1|\rho)-\mathfrak{I}(\mathcal{O}_1|\Phi_{\mathcal{O}_2}(\rho)),
\ee 
whose nonnegativity is implied by the theorem given below. Invariant under permutation of indices, as can be noted from its symmetric form $\mathcal{N}(\mathcal{O}_1,\mathcal{O}_2|\rho)=S(\Phi_{\mathcal{O}_1}(\rho))+S(\Phi_{\mathcal{O}_2}(\rho))-S(\Phi_{\mathcal{O}_1\mathcal{O}_2}(\rho))-S(\rho)$, this measure quantifies nonlocal aspects associated with the couple $\{\mathcal{O}_1,\mathcal{O}_2\}$ given $\rho$. We also define the {\em minimal nonlocality} of a preparation $\rho$ as the maximally restrictive optimization over the observables, i.e.,
\be \label{N}
\mathcal{N}_{\mathrm{min}}(\rho)\equiv \min\limits_{\text{\tiny $\{\mathcal{O}_1,\mathcal{O}_2\}$}}\mathcal{N}(\mathcal{O}_1,\mathcal{O}_2|\rho),
\ee 
whose bounds are defined by the following result. Let $\mathcal{D}_{12}(\rho)=\min_{\text{\tiny $\{\mathcal{O}_1,\mathcal{O}_2\}$}}D_{[\mathcal{O}_1,\mathcal{O}_2]}(\rho)$ be the {\em global quantum discord}~\cite{rulli11}, with $D_{[\mathcal{O}_1,\mathcal{O}_2]}(\rho)=I_{1:2}(\rho)-I_{1:2}(\Phi_{\mathcal{O}_1\mathcal{O}_2}(\rho))$.

\vskip2mm
\noindent {\bf Theorem.} {\em Given an arbitrary preparation $\rho\in\mathcal{H}_1\otimes\mathcal{H}_2$, it holds that $0\leq \mathcal{N}_{\mathrm{min}}(\rho)\leq \mathcal{D}_{12}(\rho)$. In particular, if $\rho$ is pure, then $\mathcal{N}_{\mathrm{min}}(\rho)=0$.}
\vskip2mm

%======== PROOF ============================================
\noindent {\em Proof.} From Eqs.~\eqref{IO1} and \eqref{NO1O2}, one shows that $\mathcal{N}(\mathcal{O}_1,\mathcal{O}_2|\rho)=D_{[\mathcal{O}_1]}(\rho)+D_{[\mathcal{O}_2]}(\rho)-D_{[\mathcal{O}_1,\mathcal{O}_2]}(\rho)$, where $D_{[\mathcal{O}_1,\mathcal{O}_2]}(\rho)=I_{1:2}(\rho)-I_{1:2}(\Phi_{\mathcal{O}_1\mathcal{O}_2}(\rho))$. From the non-negativity of the quantum discord $\mathcal{D}_k$, it follows that $I_{1:2}(\rho)\geq I_{1:2}(\Phi_{k}(\rho))$, $k=1,2$. With the replacement $\rho\to \Phi_{\mathcal{O}_j}(\rho)$ we get $I_{1:2}(\Phi_{j}(\rho))\geq I_{1:2}(\Phi_{\mathcal{O}_1\mathcal{O}_2}(\rho))$ and
\be 
I_{1:2}(\Phi_{1}(\rho))+ I_{1:2}(\Phi_{2}(\rho))\geq 2I_{1:2}(\Phi_{\mathcal{O}_1\mathcal{O}_2}(\rho)).\nonumber
\ee 
By rewriting $I_{1:2}$ in terms of its discordlike quantity $D$ [see the relation following Eq.~(4)] we obtain
\be
\mathcal{N}(\mathcal{O}_1,\mathcal{O}_2|\rho)=D_{[\mathcal{O}_1]}+D_{[\mathcal{O}_2]}-D_{[\mathcal{O}_1\mathcal{O}_2]}\leq D_{[\mathcal{O}_1\mathcal{O}_2]},\nonumber
\ee 
which, upon minimization, gives
\be \label{N<D}
\mathcal{N}_{\mathrm{min}}(\rho)\leq \mathcal{D}_{12}(\rho).
\ee 
This upper bound reveals that the absence of quantum discord allows for the existence of a couple of observables for which nonlocality will not manifest. This point reveals a hierarchy which is similar to that exhibited between entanglement and Bell nonlocality. 

The proof of the lower bound goes as follows. Consider a quadripartite state $\varrho_0=\rho\otimes\rho_A\otimes\rho_B$, where $\rho\in\mathcal{H}_1\otimes\mathcal{H}_2$, $\rho_A=\ket{a_0}\bra{a_0}\in\mathcal{H}_A$, and $\rho_B=\ket{b_0}\bra{b_0}\in\mathcal{H}_B$. Let $\mathcal{O}_1=\sum_ko_{1k}\mathrm{O}_{1k}$ and $\mathcal{O}_2=\sum_jo_{2j}\mathrm{O}_{1j}$ be observables acting on $\mathcal{H}_1$ and $\mathcal{H}_2$, respectively, where $\mathrm{O}_{1k}=\ket{k}\bra{k}$ and $\mathrm{O}_{2j}=\ket{j}\bra{j}$. Consider a unitary transformation $U=U_{1A}\otimes U_{2B}$ such that $\varrho=U\varrho_0 U^{\dag}$, $U_{1A}\ket{k}\ket{a_0}=\ket{k}\ket{a_k}$, and $U_{2B}\ket{j}\ket{b_0}=\ket{j}\ket{b_j}$. It follows that
\be 
\varrho\!=\!\sum\limits_{\text{\tiny $\begin{smallmatrix} k, k' \\ j,j' \end{smallmatrix}$}} \langle kj|\rho|k'j'\rangle \ket{k}\bra{k'}\otimes\ket{j}\bra{j'}\otimes\ket{a_k}\bra{a_{k'}}\otimes\ket{b_j}\bra{b_{j'}}.\nonumber 
\ee 
Now consider the following partitions: $X=A$, $Y=B$, and $Z=12$. Using the above state, one may show that 
\be
\begin{array}{lll}
S(\rho_{XYZ})=S(\rho), & \quad & S(\rho_Z)=S(\Phi_{\mathcal{O}_1\mathcal{O}_2}(\rho)),  \\ \\
S(\rho_{XZ})=S(\Phi_{\mathcal{O}_2}(\rho)), & & S(\rho_{YZ})=S(\Phi_{\mathcal{O}_1}(\rho)), 
\end{array}\nonumber
\ee
where $\rho_Z$, $\rho_{XZ}$, and $\rho_{YZ}$ are reductions of $\varrho=\rho_{XYZ}$. From the strong subaditivity of the von Neumann entropy, $S(\rho_{XYZ})+S(\rho_Z)\leq S(\rho_{XZ})+S(\rho_{YZ})$ \cite{chuang}, it follows that $\mathcal{N}(\mathcal{O}_1,\mathcal{O}_2|\rho)\geq 0$ for all $\{\mathcal{O}_1,\mathcal{O}_2\}$, so that
\be \label{N>0}
\mathcal{N}_{\mathrm{min}}(\rho)\geq 0.
\ee 
The result for pure states is proved as follows. Take observables $\mathcal{O}_1=\sum_ko_{1k}\ket{k}\bra{k}$ and $\mathcal{O}_2=\sum_jo_{2j}\ket{j}\bra{j}$ that are connected with the Schmidt decomposition $\ket{\psi}=\sum_k\sqrt{\lambda_k}\ket{k}\ket{k}$ in the following peculiar way: $\mathcal{O}_1$'s eigenstates $\{\ket{k}\}$ correspond to the Schmidt sub-basis $\{\ket{k}\}$ whereas $\mathcal{O}_2$'s eigenstates $\{\ket{j}\}$ form a MUB with the Schmidt sub-basis $\{\ket{k}\}$, i.e., $|\langle k|j\rangle|^2=\tfrac{1}{d_2}$, where $d_2=\dim(\mathcal{H}_2)$. For these observables one shows that $S(\Phi_{\mathcal{O}_2}(\ket{\psi}))=\ln d_2$ and $S(\Phi_{\mathcal{O}_1\mathcal{O}_2}(\ket{\psi}))=\ln d_2+S(\Phi_{\mathcal{O}_1}(\ket{\psi}))$, from which it immediately follows that $\mathcal{N}(\mathcal{O}_1,\mathcal{O}_2|\ket{\psi})=0$. Therefore, 
\be \label{N=0}
\mathcal{N}_{\mathrm{min}}(\ket{\psi})=0. 
\ee 
This result also follows from \eqref{N>0} and from the fact that pure states saturate the strong subaditivity (see Ref.~\cite{costa14}). Equations \eqref{N<D}-\eqref{N=0} define the content of the theorem. \hfill{\scriptsize $\blacksquare$}

\begin{figure}[ht]
\includegraphics[width=\columnwidth]{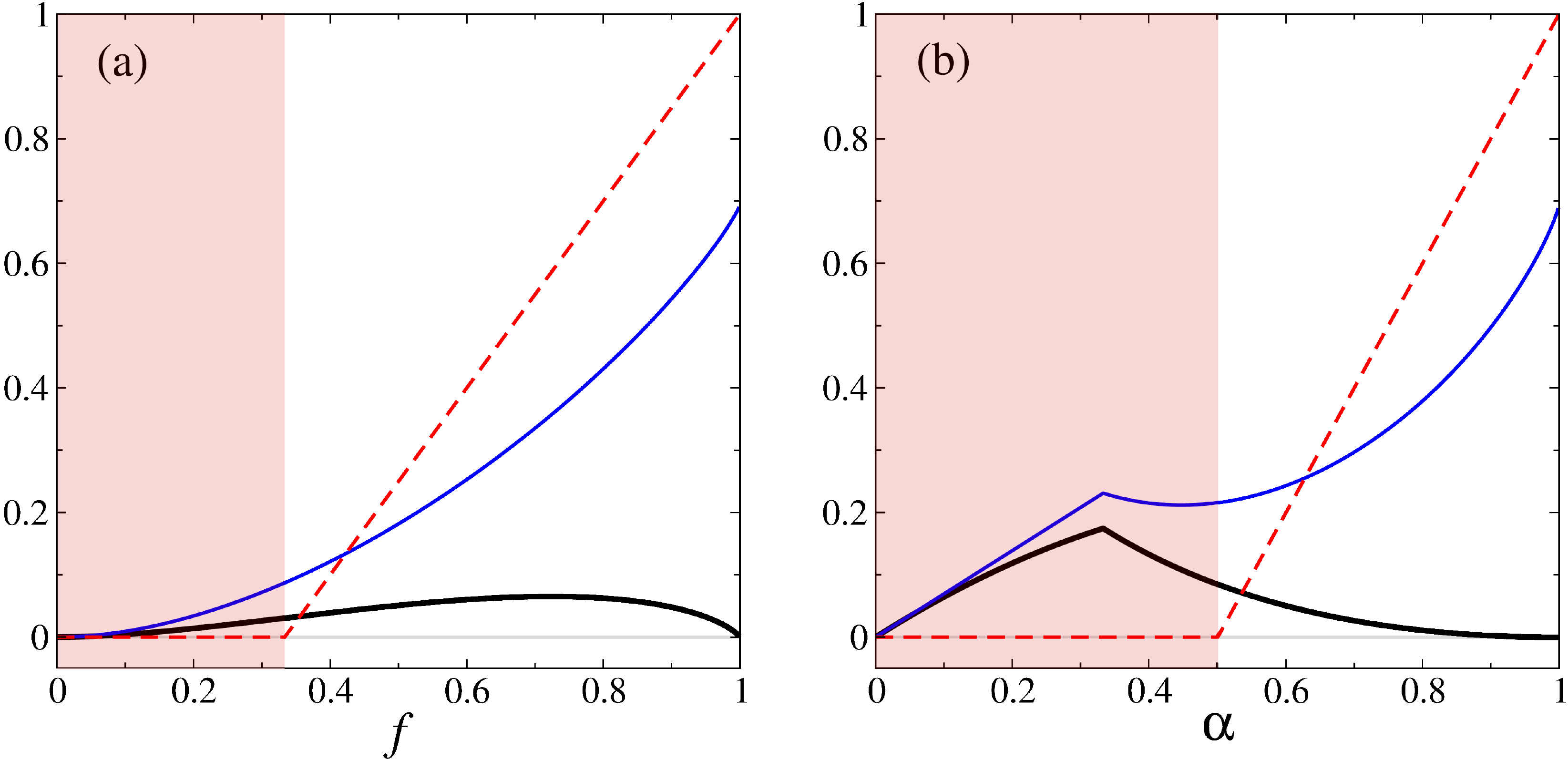}
\caption{(Color online). Minimal nonlocality $\mathcal{N}_{\mathrm{min}}$ (thick black line), global quantum discord $\mathcal{D}_{12}$ (blue line) and entanglement $E$ (dashed red line) for (a) $\rho_W$ and (b) $\rho_{\alpha}$. In the pink shaded area, $E=0$ whereas $\mathcal{N}\geq 0$ for all $\{\mathcal{O}_1,\mathcal{O}_2\}$.}
\label{fig2}
\end{figure}

It can be checked by inspection that, while $\mathcal{N}=0$ for a fully uncorrelated state $\rho=\rho_1\otimes\rho_2$, the same cannot be ensured for a separable state $\rho=\sum_kp_k\rho_{1k}\otimes\rho_{2k}$. To illustrate this point, let us compute $\mathcal{N}_{\mathrm{min}}$ for specific two-qubit preparations, namely, the Werner state $\rho_W=\tfrac{(1-f)}{4}\mathbbm{1}\otimes\mathbbm{1}+f\ket{s}\bra{s}$, where $f$ measures the fidelity of $\rho_W$ with the singlet $\ket{s}=\tfrac{1}{\sqrt{2}}\big(\ket{+r}\ket{-r}-\ket{-r}\ket{+r}\big)$ ($r=x,y,z$), and the $\alpha$-state, $\rho_{\alpha}=\tfrac{\mathbbm{1}\otimes\mathbbm{1}}{4}+\tfrac{\alpha}{4}(\sigma_1\otimes\sigma_1-\sigma_2\otimes\sigma_2)+\tfrac{2\alpha-1}{4}\sigma_3\otimes\sigma_3$. Although analytical results have been computed, they are not insightful and will be omitted. They are shown in Fig.~\ref{fig2} along with the results for the global quantum discord $\mathcal{D}_{12}$ and the entanglement $E$ (as quantified by concurrence). Two aspects are remarkable. First, unlike Bell nonlocality, $\mathcal{N}$ may exist even in the absence of entanglement (see Refs.~\cite{bennett99,walgate02,luo11,zhang14,modi12} for other examples of nonlocality without entanglement and a link with discord). This suggests an interesting analogy: $\mathcal{N}$ is able to capture nonlocal aspects to which Bell inequalities are insensitive just like quantum discord can detect correlations that are invisible to entanglement measures. Second, $\mathcal{N}_{\mathrm{min}}$ vanishes for pure states ($f=\alpha=1$), as anticipated by the Theorem. This does not mean that pure states prevent nonlocality to manifest, but that there exists al least a couple of observables for which $\mathcal{N}$ vanishes (see the theorem proof). In fact, it is not difficult to show that, if we take observables $\mathcal{O}_1$ and $\mathcal{O}_2$ whose eigenstates define the Schmidt basis, then  $\mathcal{N}(\mathcal{O}_1,\mathcal{O}_2|\ket{\psi})=E(\ket{\psi})$, with $E(\ket{\psi})=-\text{Tr}_1(\rho_1\ln\rho_1)$ and $\rho_1=\text{Tr}_2\ket{\psi}\bra{\psi}$, gives the entanglement entropy of $\ket{\psi}$.  The take-away message here comes as follows. Take the singlet $\rho_s=\ket{s}\bra{s}$ as an example. By direct calculation we can check that $\mathcal{N}(\sigma_{1r},\sigma_{2r'}|\rho_s)=\delta_{r,r'}E(\ket{s})$, where $r,r'=x,y,z$ and $\delta_{r,r'}$ is the Kronecker delta. This shows that nonlocal aspects appear when we look at the couple of observables in which the entanglement has been encoded, namely, the observables that define the Schmidt basis. No nonlocality is detected when we look at a couple of MIO, one of them composing the Schmidt basis. This explains why $\mathcal{N}_{\mathrm{min}}(\ket{\psi})=0$ and stress that $\mathcal{N}$ is conceptually different from quantum discord, which reduces to the entanglement entropy for pure states.
%

%======================================================
{\em Nonlocality from irreality and measurement}. When a correlation is created via physical interactions in a closed quantum system, a {\em constraint} is established which manifests as a conservation law. For instance, for the singlet $\ket{s}=\ket{S=0,M=0}$ one has that $\mathcal{S}_{z}=\mathcal{S}_{1z}+\mathcal{S}_{2z}=0$. Here the situation is such that even though the total $z$-spin $\mathcal{S}_z$ is real (i.e., fully definite, as ensured by the physical interaction) the individuals $\mathcal{S}_{1z}$ and $\mathcal{S}_{2z}$ are not. The crux is that the constraint $\mathcal{S}_{2z}=-\mathcal{S}_{1z}$ reduces the {\em a priori} independent indefiniteness of $\mathcal{S}_{1z}$ and $\mathcal{S}_{2z}$ to a state of {\em conditional reality}, a situation in which the reality of an observable becomes conditioned to the reality of another. Once one of them gets real, so does the other. By separating the subsystems without degrading the constraint, one will be able to define the reality of an observable by means of an action in a remote site. This does not occur classically, because even though the notion of a nonlocally spread constraint keeps valid, there is no fundamental indefiniteness underlying the observables, i.e., their realities are already established before the separation. It is immediately seen, therefore, that irreality is a basic condition for the manifestation of quantum nonlocality. Indeed, as is predicted by $\mathcal{N}(\mathcal{O}_1,\mathcal{O}_2|\Phi_{\mathcal{O}_k}(\rho))=0$ $(k=1,2)$, there is no nonlocality for a preparation in which one of the observables is already real.

There is, however, a second condition for nonolocality to manifest: the reality in a given site has to be established to promote the aforementioned conditionalization in the remote site. This is precisely the point behind the conception of Eq.~\eqref{NO1UA}, where the generation of correlations and the posterior discard of the ancilla played the role an unread measurement. In fact, when we are able to access the whole system, including the ancilla, then the reality of the subsystem does not get defined, as we have seen in Bohr's floating slit thought experiment. This point can be further illustrated with some generality as follows. Consider a preparation $\varrho$ for a multipartite system and an arbitrary partition $\mathcal{H}_x\otimes\mathcal{H}_y$ for the Hilbert space. Let $U_y$ be a unitary transformation in $\mathcal{H}_y$ and $\mathcal{O}_1\in\mathcal{H}_1\in\mathcal{H}_x$. Given that $\Phi_{\mathcal{O}_1}$ and $U_y$ commute, it follows that
\be \label{noNL}
\mathfrak{I}(\mathcal{O}_1|\varrho)-\mathfrak{I}(\mathcal{O}_1|U_y\varrho U_y^{\dag})=0.
\ee 
We see that the reality of a given observable can never be changed by {\em unitary} actions occurring in remote parts of the system. This result shows that, if we could access the whole system, no nonlocality would be detected in nature. (Recently, a similar conclusion has been reached, via arguments grounded on the framework of the many-worlds interpretation~\cite{tipler14}.) However, physics is fundamentally concerned with experiments, where discarding a system---a nonunitary operation---is an irremediable fact. Indeed, in a measurement process, the degrees of freedom of the probed system always remains veiled to the observer; this is why we have to use an accessible pointer in the first place.

%===================
{\em Conclusion}. Based on the premise that an element of reality exists for an observable that has been measured, and using a protocol involving two observers, we developed a quantifier that extends current views of reality in two ways. First, our measure moves the focus from the quantum state to the couple state-observable. Second, it is able to diagnose reality also when mixed states are concerned. In particular, our approach allows for the identification of epistemic states in quantum theory. Once one accepts our proposition, the following framework is implied. i)~Upon discarding the system of interest, the apparatus is left in a state for which the pointer is real, so that the state reduction can be viewed as mere information updating. ii)~Noncommuting observables cannot be simultaneously real for entangled states, this being a result that is in contrast with EPR's claim. iii)~Quantum correlations forbid the concept of a separable reality, even when the subsystems are arbitrarily far apart from each other. iv)~Alterations induced in the reality of a given observable by means of measurements performed in a remote system reveal nonlocal aspects which are not captured by Bell inequality violations. Finally, it is worth noting that many relevant questions concerning our notions of reality and nonlocality can be formulated in the contexts of multipartite systems, quantum reference frames, weak measurements, and thermodynamics~\cite{comment}. This paves the way for an interesting research program.

Put in the perspective of EPR's work, our results propose a scenario in which quantum mechanics can be viewed as a complete theory, provided we accept that observables can be fundamentally indefinite and nonlocality is unavoidable. To our perception, these aspects are by now viewed as established facts by a significant part of the scientific community, for which our measures then emerge as relevant tools.

%===============
\acknowledgments
This work was supported by CNPq/Brazil and the National Institute for Science and Technology of Quantum Information (INCT-IQ/CNPq). The authors acknowledge A. D. Ribeiro, F. Parisio, and M. S. Sarandy for discussions. 

%==========================

%==========================================================

\begin{thebibliography}{99}
\bibitem{comment} In particular, employing the relation $W(\rho)=k_BTI(\rho)$ \cite{zurek03,horodecki05}, where $I(\rho)=\ln{d}-S(\rho)$, $\rho\in\mathcal{H}$, $d=\dim{\mathcal{H}}$, and $k_B$ is the Boltzmann constant, one may follow the proposals of Refs.~\cite{ana13,angelo15,zurek03} to conceive an operational interpretation for $\mathfrak{I}(\mathcal{O}_1|\rho)$ in terms of the thermodynamic work $W(\rho)$ that can be extracted from a reservoir at temperature $T$. For a single system, one can readily write $k_BT\mathfrak{I}(\mathcal{O}|\rho)=W(\rho)-W(\Phi_{\mathcal{O}}(\rho))$, which shows that the irreality of $\mathcal{O}$ given the preparation $\rho$ provides thermodynamic advantage in relation to a preparation $\Phi_{\mathcal{O}}(\rho)$ for which $\mathcal{O}$ is real. Research along this line is now in progress.
\end{thebibliography}
\end{document}